\documentclass{elsarticle}
\usepackage{hyperref}
\usepackage{amsmath}
\usepackage{graphicx}
\graphicspath{{figures/}}

\usepackage{color}

\newcommand{\MSbar}{\ensuremath{\overline{\text{MS}}}}

\newenvironment{absolutelynopagebreak}
  {\par\nobreak\vfil\penalty0\vfilneg
   \vtop\bgroup}
  {\par\xdef\tpd{\the\prevdepth}\egroup
   \prevdepth=\tpd}

\begin{document}
\begin{absolutelynopagebreak}
  \begin{frontmatter}
    \begin{flushright}
      DESY-15-091\\
      TUM-HEP-999/15
    \end{flushright}
    \title{Low-energy moments of non-diagonal quark current correlators
      at four loops} \author[Munich]{A.~Maier}
    \author[Zeuthen]{P.~Marquard} \address[Munich]{ Physik Department
      T31, James-Franck-Stra\ss{}e, Technische Universit\"at M\"unchen,
      D-85748 Garching, Germany } \address[Zeuthen]{Deutsches Elektronen
      Synchrotron DESY, Platanenallee 6, D-15738 Zeuthen, Germany}
    \begin{abstract}
      We compute the leading four physical terms in the low-energy
      expansions of heavy-light quark current correlators at four-loop
      order. As a by-product we reproduce the corresponding top-induced
      non-singlet correction to the electroweak $\rho$ parameter.
    \end{abstract}
  \end{frontmatter}
\end{absolutelynopagebreak}
\section{Introduction}
\label{sec:introduction}

Two-point correlation functions of heavy-light quark currents have found
use in a number of phenomenological applications. One example is the
prediction of corrections to the electroweak $\rho$
parameter~\cite{Schroder:2005db,Chetyrkin:2006bj,Boughezal:2006xk},
where the flavour non-diagonal correlator of vector currents is required
for vanishing external momentum. Another important class of applications
is the sum-rule determination of meson decay constants (see
e.g.~\cite{Penin:2001ux,Jamin:2001fw}). Here, the absorptive part of the
respective correlators above the production threshold is
needed.

Progress in lattice simulation may allow precise determinations of even
more QCD parameters. For instance, the values of the strong coupling
constant, the charm quark mass and the bottom quark mass have been
determined with high accuracy from moments of heavy-heavy correlators
in~\cite{Allison:2008xk,McNeile:2010ji}. In these analyses, moments of
flavour diagonal currents have been determined on the lattice choosing a
frame where the spatial momentum of the correlators vanishes. The values
of the quark masses and the coupling constant are then extracted by
equating these moments to their counterparts calculated in perturbation
theory at the four-loop
order~\cite{Chetyrkin:2006xg,Boughezal:2006px,Sturm:2008eb,Maier:2008he,Hoang:2008qy,Maier:2009fz,Kiyo:2009gb}.

The methodology is thus similar to traditional quarkonium sum
rules~\cite{Novikov:1976tn,Novikov:1977dq,Chetyrkin:2009fv}, but using
lattice moments in place of moments of the experimentally measured
hadronic $R$ ratio. While for the sum rules only the correlator of
vector currents can be used, there is no such restriction for the
lattice simulation. In fact, in~\cite{Allison:2008xk} different Lorentz
structures were considered, with the most precise results stemming from
pseudoscalar currents. Furthermore, also correlators of heavy-light
currents could be used to extract the values of the charm and bottom
quark masses and possibly the strong coupling
constant~\cite{Koponen:2010jy}. To be competitive with the analyses for
the heavy-heavy case the corrections to the perturbative moments of the
heavy-light current correlators have to be known up to four loops. These
corrections are presented in this work.

Given their usefulness, perturbative corrections to heavy-light
correlators have been studied quite intensively and analytic results up
to two loops have been known for many
years~\cite{Djouadi:1993ss,Djouadi:1994gf}.  While the three-loop
correction is not known analytically, many terms in expansions in both
the low-energy and the high-energy limit have been calculated
in~\cite{Chetyrkin:2000mq,Chetyrkin:2001je,Maier:2011jd}. Combining
these with the behaviour near threshold, accurate approximations for
arbitrary kinematics have been
constructed~\cite{Chetyrkin:2000mq,Chetyrkin:2001je}.  In the low-energy
region also corrections due to a non-vanishing light quark mass are
known~\cite{Hoff:2011ge,Grigo:2012ji}.

The four-loop corrections remain mostly unknown. In the high-energy
region the leading term is equal to the non-singlet part of the
corresponding diagonal correlator, which has been computed for both
scalar and vector
currents~\cite{Baikov:2004ku,Baikov:2009uw,Chetyrkin:2010dx}. In the
low-energy region, conversely, there is no such simple correspondence
between diagonal and non-diagonal correlators. The vector correlator in
the limit of vanishing external momentum constitutes a central
ingredient in the determination of non-singlet four-loop corrections to
the $\rho$ parameter, which have been calculated
in~\cite{Chetyrkin:2006bj,Boughezal:2006xk}.

In this work we present the four-loop corrections to the low-energy
expansions of both scalar and vector heavy-light quark current
correlators up to the eighth power of the external momentum. After
introducing our conventions in section~\ref{sec:conventions}, we briefly
describe the calculational setup and present our results in
section~\ref{sec:calculation}. Section~\ref{sec:rho} describes the
re-calculation of the top-induced contributions to the electroweak
$\rho$ parameter, which constitutes an important consistency check. We
conclude in section~\ref{sec:conclusion}.

\section{Conventions}
\label{sec:conventions}

The correlators of heavy-light vector and scalar currents are defined as
\begin{align}
  \label{eq:corr_v}
  \Pi_{\mu\nu}(q) ={}& i \int dx \, e^{iqx}\langle 0 | j_\mu(x) j_\nu(0) |
  0\rangle\,,\\
  \label{eq:corr_s}
  \Pi(q) ={}& i \int dx \, e^{iqx}\langle 0 | j(x) j(0) | 0\rangle
\end{align}
with the vector current $j_\mu(x) = \bar{\psi}(x) \gamma_\mu \chi(0)$
and the scalar current $j(x) = \bar{\psi}(x) \chi(0)$. We consider a
heavy quark $\psi$ with the pole mass $m$ and a massless light quark
$\chi$. It should be noted that in the limit of a vanishing
light-quark mass the correlators of two axial-vector or pseudo-scalar
currents coincide with the vector and scalar correlators, respectively.

It is convenient to introduce polarisation functions
\begin{align}
  \label{eq:Pi_v_def}
  \Pi_{\mu\nu}(q) ={}& (-q^2g_{\mu\nu} + q_\mu q_\nu)\Pi^v(q^2) + q_\mu
  q_\nu \Pi^v_L(q^2)\,,\\
  \label{eq:Pi_s_def}
  \Pi(q) ={}& q^2 \Pi^s(q^2)\,.
\end{align}
In the following we will not consider the longitudinal polarisation
$\Pi^v_L(q^2)$. The perturbative expansions of $\Pi^\delta(q^2)$ with
$\delta = v,s$ up to four loops read
\begin{equation}
  \label{eq:Pi_pert}
  \Pi^\delta(q^2) = \Pi^{\delta,(0)}(q^2)
  + \frac{\alpha_s}{\pi}C_F\Pi^{\delta,(1)}(q^2)
  + \bigg(\frac{\alpha_s}{\pi}\bigg)^2\Pi^{\delta,(2)}(q^2)
  + \bigg(\frac{\alpha_s}{\pi}\bigg)^3\Pi^{\delta,(3)}(q^2)
  + \dots\,.
\end{equation}
Being interested in the limit $q^2\to 0$, we can expand the coefficients
in the above series as
\begin{equation}
  \label{eq:C_def}
  \Pi^{\delta,(i)}(q^2) =  \frac{3}{16 \pi^2}\sum_{n=-1}^\infty C^{\delta,(i)}_n z^n = \frac{3}{16 \pi^2}\sum_{n=-1}^\infty \bar{C}^{\delta,(i)}_n \bar{z}^n\,,
\end{equation}
where we have used the abbreviations $z = q^2/m^2, \bar{z} =
q^2/\bar{m}^2$ with $\bar{m}$ denoting the mass of the heavy quark in the
\MSbar{} scheme. Note that the coefficients with $n=-1, 0$ still contain
poles in the limit $\epsilon = (4-d)/2 \to 0$. In physical observables
these have to be cancelled by the wave-function and mass renormalisations
of the particles (e.g. $W$ bosons) coupling to the respective
current. In the following we will describe the calculation of the
coefficients $C^{\delta,(3)}_n, \bar{C}^{\delta,(3)}_n$ for $n \leq 4$.

\section{Calculation and results}
\label{sec:calculation}

First, the four-loop diagrams contributing to the polarisation functions
are generated with \texttt{QGRAF}~\cite{Nogueira:1991ex}. In the
following steps we perform several algebraic manipulations with the help
of \texttt{TFORM}~\cite{Vermaseren:2000nd,Tentyukov:2007mu}. As a first
simplification, we apply partial fractioning to denominators that differ
only by their mass and the external momentum $q$, i.e. we use
\begin{equation}
  \label{eq:part_frac}
  \frac{1}{p^2}\ \frac{1}{(p\pm q)^2 - m^2} = \frac{1}{q^2\pm 2pq-m^2}\bigg(\frac{1}{p^2} - \frac{1}{(p\pm q)^2-m^2}\bigg)\,.
\end{equation}
Since we will perform an expansion in $q$ the prefactor on the
right-hand side has no influence on the tadpole topology of the
considered diagram. Performing partial fractioning before the
identification of the diagram topologies greatly reduces both the number
and the complexity of the topologies that have to be considered. Using
the algorithm described in
\ref{sec:symmetrisation} we map the
resulting diagrams onto 28 topologies.

Next, colour factors are calculated using the
\texttt{FORM}~\cite{Vermaseren:2000nd} package
\texttt{color}~\cite{vanRitbergen:1998pn}.  We choose a routing
  for the external momentum $q$ which minimizes the number of
  propagators depending on $q$. After this we evaluate the traces
over gamma matrices and perform a Taylor expansion in $q$. The scalar
integrals we obtain after tensor reduction and the elimination of
reducible scalar products are reduced to master integrals using a
private implementation\footnote{ The implementation is written in C++
and uses~\texttt{GiNaC}~\cite{Bauer:2000cp} and
\texttt{fermat}~\cite{fermat}.}~\cite{crusher} of Laporta's
algorithm~\cite{Laporta:2001dd}. All required master integrals are
known analytically or
numerically~\cite{Schroder:2005va,Faisst:2006sr,Maierhoefer:2006}.

For the presentation of our results we impose the overall
renormalisation condition $\Pi^v(0) = \Pi^s(0) = 0$. The corresponding
divergent subtraction terms are listed in
\ref{sec:subtr}. For the remaining coefficients according to
equation~(\ref{eq:C_def}) we obtain\footnote{The moments are presented
in a form that yields five significant digits for QCD with $n_l \leq 5$
light flavours. All results are attached in electronic form as ancillary
files to this preprint.}
\begin{align}
\label{eq:C_v_1_OS}
C^{v,(3)}_1 ={}&+14.5508\,\*C_A^2\*C_F + 8.4892\,\*C_A\*C_F^2 + 0.351\,\*C_F^3 \notag\\
&- 0.2294\,\*C_A\*C_F\*T_F\*n_h - 0.6242\,\*C_F^2\*T_F\*n_h \notag\\
&- 12.56835\,\*C_A\*C_F\*T_F\*n_l - 3.07525\,\*C_F^2\*T_F\*n_l \notag\\
&+ 0.107\,\*C_F\*T_F^2\*n_h^2 + 0.14\,\*C_F\*T_F^2\*n_h\*n_l + 1.91917\,\*C_F\*T_F^2\*n_l^2\,,
\displaybreak[0]\\
\label{eq:C_v_2_OS}
C^{v,(3)}_2 ={}&+7.39116\,\*C_A^2\*C_F + 5.65943\,\*C_A\*C_F^2 + 0.80504\,\*C_F^3 \notag\\
&+ 0.0683\,\*C_A\*C_F\*T_F\*n_h - 0.3114\,\*C_F^2\*T_F\*n_h \notag\\
&- 6.0806\,\*C_A\*C_F\*T_F\*n_l - 2.2303\,\*C_F^2\*T_F\*n_l \notag\\
&+ 0.008\,\*C_F\*T_F^2\*n_h^2 - 0.0052\,\*C_F\*T_F^2\*n_h\*n_l + 0.9442\,\*C_F\*T_F^2\*n_l^2\,,
\displaybreak[0]\\
\label{eq:C_v_3_OS}
C^{v,(3)}_3 ={}&+4.42563\,\*C_A^2\*C_F + 3.86666\,\*C_A\*C_F^2 + 0.73105\,\*C_F^3 \notag\\
&+ 0.0448\,\*C_A\*C_F\*T_F\*n_h - 0.1713\,\*C_F^2\*T_F\*n_h \notag\\
&- 3.57037\,\*C_A\*C_F\*T_F\*n_l - 1.5461\,\*C_F^2\*T_F\*n_l \notag\\
&+ 0.0017\,\*C_F\*T_F^2\*n_h^2 - 0.005\,\*C_F\*T_F^2\*n_h\*n_l + 0.56396\,\*C_F\*T_F^2\*n_l^2\,,
\displaybreak[0]\\
\label{eq:C_v_4_OS}
C^{v,(3)}_4 ={}&+2.90512\,\*C_A^2\*C_F + 2.7515\,\*C_A\*C_F^2 + 0.5965\,\*C_F^3 \notag\\
&+ 0.0278\,\*C_A\*C_F\*T_F\*n_h - 0.104\,\*C_F^2\*T_F\*n_h \notag\\
&- 2.31867\,\*C_A\*C_F\*T_F\*n_l - 1.10502\,\*C_F^2\*T_F\*n_l \notag\\
&+ 0.0006\,\*C_F\*T_F^2\*n_h^2 - 0.003\,\*C_F\*T_F^2\*n_h\*n_l + 0.3708\,\*C_F\*T_F^2\*n_l^2\,,
\displaybreak[0]\\
\label{eq:C_s_1_OS}
C^{s,(3)}_1 ={}&+1.6424\,\*C_A^2\*C_F + 1.65318\,\*C_A\*C_F^2 + 1.41042\,\*C_F^3 \notag\\
&- 1.39916\,\*C_A\*C_F\*T_F\*n_h + 0.5551\,\*C_F^2\*T_F\*n_h \notag\\
&- 3.129699\,\*C_A\*C_F\*T_F\*n_l + 0.556834\,\*C_F^2\*T_F\*n_l \notag\\
&+ 0.376\,\*C_F\*T_F^2\*n_h^2 + 0.65308\,\*C_F\*T_F^2\*n_h\*n_l + 0.441495\,\*C_F\*T_F^2\*n_l^2\,,
\displaybreak[0]\\
\label{eq:C_s_2_OS}
C^{s,(3)}_2 ={}&+5.66925\,\*C_A^2\*C_F + 5.36995\,\*C_A\*C_F^2 + 2.1099\,\*C_F^3 \notag\\
&- 0.0476\,\*C_A\*C_F\*T_F\*n_h + 0.1338\,\*C_F^2\*T_F\*n_h \notag\\
&- 5.00716\,\*C_A\*C_F\*T_F\*n_l - 1.60465\,\*C_F^2\*T_F\*n_l \notag\\
&+ 0.037\,\*C_F\*T_F^2\*n_h^2 + 0.0314\,\*C_F\*T_F^2\*n_h\*n_l + 0.711301\,\*C_F\*T_F^2\*n_l^2\,,
\displaybreak[0]\\
\label{eq:C_s_3_OS}
C^{s,(3)}_3 ={}&+4.18695\,\*C_A^2\*C_F + 4.9201\,\*C_A\*C_F^2 + 2.0783\,\*C_F^3 \notag\\
&+ 0.0196\,\*C_A\*C_F\*T_F\*n_h - 0.0103\,\*C_F^2\*T_F\*n_h \notag\\
&- 3.5077\,\*C_A\*C_F\*T_F\*n_l - 1.7089\,\*C_F^2\*T_F\*n_l \notag\\
&+ 0.009\,\*C_F\*T_F^2\*n_h^2 + 0.0006\,\*C_F\*T_F^2\*n_h\*n_l + 0.5215\,\*C_F\*T_F^2\*n_l^2\,,
\displaybreak[0]\\
\label{eq:C_s_4_OS}
C^{s,(3)}_4 ={}&+2.945114\,\*C_A^2\*C_F + 3.81782\,\*C_A\*C_F^2 + 1.6905\,\*C_F^3 \notag\\
&+ 0.01994\,\*C_A\*C_F\*T_F\*n_h - 0.0347\,\*C_F^2\*T_F\*n_h \notag\\
&- 2.41729\,\*C_A\*C_F\*T_F\*n_l - 1.3927\,\*C_F^2\*T_F\*n_l \notag\\
&+ 0.0034\,\*C_F\*T_F^2\*n_h^2 - 0.0016\,\*C_F\*T_F^2\*n_h\*n_l + 0.36976\,\*C_F\*T_F^2\*n_l^2\,,
\end{align}
where we have set the renormalisation scale $\mu$ to the on-shell mass $m$. We follow the
usual convention for the colour factors with $C_A = 3, C_F = 4/3, T_f =
1/2$ for QCD. The number of light (massless) quark flavours is denoted
by $n_l$, whereas $n_h$ stands for the number of heavy flavours.

If we choose to express the polarisation functions in terms of the
\MSbar{} mass $\bar{m}$ at the scale $\mu = \bar{m}$ and
$\alpha_s(\bar{m})$, we arrive at
\begin{align}
\label{eq:C_v_1_MS}
\bar{C}^{v,(3)}_1 ={}&-1.2994791\,\*C_A^2\*C_F + 1.20957\,\*C_A\*C_F^2 + 0.537098\,\*C_F^3 \notag\\
&- 1.75125\,\*C_A\*C_F\*T_F\*n_h + 1.29201\,\*C_F^2\*T_F\*n_h \notag\\
&+ 0.530618\,\*C_A\*C_F\*T_F\*n_l - 0.0193\,\*C_F^2\*T_F\*n_l \notag\\
&- 0.0853\,\*C_F\*T_F^2\*n_h^2 + 0.07322\,\*C_F\*T_F^2\*n_h\*n_l - 0.0389\,\*C_F\*T_F^2\*n_l^2\,,
\displaybreak[0]\\
\label{eq:C_v_2_MS}
\bar{C}^{v,(3)}_2 ={}&-1.0623284\,\*C_A^2\*C_F + 1.035507\,\*C_A\*C_F^2 + 0.1608\,\*C_F^3 \notag\\
&- 0.74336\,\*C_A\*C_F\*T_F\*n_h + 0.72663\,\*C_F^2\*T_F\*n_h \notag\\
&+ 0.905515\,\*C_A\*C_F\*T_F\*n_l - 0.46186\,\*C_F^2\*T_F\*n_l \notag\\
&- 0.0944796\,\*C_F\*T_F^2\*n_h^2 - 0.04078\,\*C_F\*T_F^2\*n_h\*n_l - 0.10011\,\*C_F\*T_F^2\*n_l^2\,,
\displaybreak[0]\\
\label{eq:C_v_3_MS}
\bar{C}^{v,(3)}_3 ={}&-0.8577978\,\*C_A^2\*C_F + 1.1607802\,\*C_A\*C_F^2 - 0.1496539\,\*C_F^3 \notag\\
&- 0.4624948\,\*C_A\*C_F\*T_F\*n_h + 0.520167\,\*C_F^2\*T_F\*n_h \notag\\
&+ 0.7959545\,\*C_A\*C_F\*T_F\*n_l - 0.631599\,\*C_F^2\*T_F\*n_l \notag\\
&- 0.06236\,\*C_F\*T_F^2\*n_h^2 - 0.027228\,\*C_F\*T_F^2\*n_h\*n_l - 0.08873276\,\*C_F\*T_F^2\*n_l^2\,,
\displaybreak[0]\\
\label{eq:C_v_4_MS}
\bar{C}^{v,(3)}_4 ={}&-0.717803\,\*C_A^2\*C_F + 1.286187\,\*C_A\*C_F^2 - 0.40493\,\*C_F^3 \notag\\
&- 0.320038\,\*C_A\*C_F\*T_F\*n_h + 0.41065\,\*C_F^2\*T_F\*n_h \notag\\
&+ 0.67538025\,\*C_A\*C_F\*T_F\*n_l - 0.72124\,\*C_F^2\*T_F\*n_l \notag\\
&- 0.04337\,\*C_F\*T_F^2\*n_h^2 - 0.018157\,\*C_F\*T_F^2\*n_h\*n_l - 0.076781\,\*C_F\*T_F^2\*n_l^2\,,
\displaybreak[0]\\
\label{eq:C_s_1_MS}
\bar{C}^{s,(3)}_1 ={}&+1.642401\,\*C_A^2\*C_F - 0.510074\,\*C_A\*C_F^2 + 1.41042\,\*C_F^3 \notag\\
&- 1.39916\,\*C_A\*C_F\*T_F\*n_h + 1.34177\,\*C_F^2\*T_F\*n_h \notag\\
&- 3.1297\,\*C_A\*C_F\*T_F\*n_l + 1.343472\,\*C_F^2\*T_F\*n_l \notag\\
&+ 0.3759\,\*C_F\*T_F^2\*n_h^2 + 0.65308\,\*C_F\*T_F^2\*n_h\*n_l + 0.4415\,\*C_F\*T_F^2\*n_l^2\,,
\displaybreak[0]\\
\label{eq:C_s_2_MS}
\bar{C}^{s,(3)}_2 ={}&+0.38582\,\*C_A^2\*C_F + 0.31725\,\*C_A\*C_F^2 + 0.916757\,\*C_F^3 \notag\\
&- 0.55487\,\*C_A\*C_F\*T_F\*n_h + 0.80004\,\*C_F^2\*T_F\*n_h \notag\\
&- 0.640838\,\*C_A\*C_F\*T_F\*n_l + 0.44938\,\*C_F^2\*T_F\*n_l \notag\\
&- 0.02698\,\*C_F\*T_F^2\*n_h^2 + 0.0092\,\*C_F\*T_F^2\*n_h\*n_l + 0.05861\,\*C_F\*T_F^2\*n_l^2\,,
\displaybreak[0]\\
\label{eq:C_s_3_MS}
\bar{C}^{s,(3)}_3 ={}&-0.039796\,\*C_A^2\*C_F + 0.48671411\,\*C_A\*C_F^2 + 0.4018122\,\*C_F^3 \notag\\
&- 0.38621\,\*C_A\*C_F\*T_F\*n_h + 0.46308\,\*C_F^2\*T_F\*n_h \notag\\
&- 0.014653\,\*C_A\*C_F\*T_F\*n_l + 0.042405\,\*C_F^2\*T_F\*n_l \notag\\
&- 0.0422\,\*C_F\*T_F^2\*n_h^2 - 0.01723\,\*C_F\*T_F^2\*n_h\*n_l - 0.00067\,\*C_F\*T_F^2\*n_l^2\,,
\displaybreak[0]\\
\label{eq:C_s_4_MS}
\bar{C}^{s,(3)}_4 ={}&-0.224943\,\*C_A^2\*C_F + 0.565757\,\*C_A\*C_F^2 + 0.067439\,\*C_F^3 \notag\\
&- 0.284427\,\*C_A\*C_F\*T_F\*n_h + 0.3319473\,\*C_F^2\*T_F\*n_h \notag\\
&+ 0.2025025\,\*C_A\*C_F\*T_F\*n_l - 0.162445\,\*C_F^2\*T_F\*n_l \notag\\
&- 0.035056\,\*C_F\*T_F^2\*n_h^2 - 0.014915\,\*C_F\*T_F^2\*n_h\*n_l - 0.021858\,\*C_F\*T_F^2\*n_l^2\,.
\end{align}

\section{The \texorpdfstring{$\mathbf{\rho}$}{rho} parameter}
\label{sec:rho}

To verify the correctness of our calculation we have performed a number
of cross checks. Obviously, our results are UV-finite. We have also
performed an expansion up to linear order in the gauge parameter and
verified that the gauge dependence cancels in the coefficients
$\bar{C}^{v,(3)}_1, \bar{C}^{s,(3)}_1$. The strongest check, however, is
the comparison to the known four-loop non-singlet corrections to the
$\rho$ parameter~\cite{Chetyrkin:2006bj,Boughezal:2006xk}.

The electroweak $\rho$ parameter has been introduced in
Ref.~\cite{Veltman:1977kh}.  Considering only QCD corrections it can be
written as
\begin{equation}
  \label{eq:1}
  \rho = 1 + \delta \rho
\end{equation}
with
\begin{equation}
  \label{eq:rhodef}
  \delta \rho = \frac{\Pi_{ZZ}(0)}{M_Z^2} -  \frac{\Pi_{WW}(0)}{M_W^2} \,,
\end{equation}
where $\Pi_{ZZ}(0)$ and $\Pi_{WW}(0)$ denote the self energies of $Z$
and $W$ boson, respectively.

In order to calculate the contribution from the $Z$-boson self energy to the $\rho$
parameter we also need the leading moment of the flavour diagonal
correlator.
To this end we introduce $\Pi^{a}_\mathrm{diag}(q^2)$ similar to
Eq.~(\ref{eq:corr_v}) but with the heavy-heavy axial current
\begin{equation}
  \label{eq:4}
  \tilde j_a^\mu = \bar \psi \gamma_5 \gamma^\mu \psi\,,
\end{equation}
and the moments
\begin{equation}
  \Pi^{a,(3)}_{\text{diag}}(q^2) =  \frac{3}{16 \pi^2}\sum_{n=-1}^\infty C^{a,(3)}_n \bigg(\frac{q^2}{4m^2}\bigg)^n\,.
\end{equation}
In what follows we will only consider the top-induced four-loop
correction to $\rho$, corresponding to $\rho_3$ in the expansion
\begin{align}
  \label{eq:rho_pert}
  \delta\rho = 3\*x_t
  \*\sum_{i=0}^{\infty}\bigg(\frac{\alpha_s}{\pi}\bigg)^i\rho_i\,, &&
  x_t = \frac{\sqrt{2}G_Fm_t^2}{16\pi^2}\,.
\end{align}
 The corresponding corrections to the $Z$
and $W$ self energies then read
\begin{align}
  \label{eq:2}
 \frac{\Pi^{(3)}_{ZZ}(0)}{M_Z^2}  ={}& 3\*x_t\left[\left
     (1- \frac{1}{d}
      \right ) C^{a,(3)}_{-1,\text{diag}}-\frac{1}{d}
   C^{a,(3)}_{L,-1,\text{diag}} \right ] \notag\\
& + \mbox{singlet terms}\,,
\end{align}
and
\begin{equation}
  \label{eq:3}
 \frac{\Pi^{(3)}_{WW}(0)}{M_W^2}  = 3\*x_t \left [\left ( 1 - \frac{1}{d}  \right ) C^{v,(3)}_{-1} -
 \frac{1}{d} C^{v,(3)}_{L,-1} \right ]\,,
\end{equation}
where the higher-order corrections $\Pi^{(3)}_{ZZ}, \Pi^{(3)}_{WW}$ are
defined in analogy to equation~(\ref{eq:Pi_pert}). $C^{v,(3)}_{L,-1}$
and $C^{a,(3)}_{L,-1,\text{diag}}$ denote
the moments with $n=-1$ of the respective longitudinal polarisation
functions;  from an explicit calculation we obtain
\begin{align}
  \label{eq:C_long}
  C^{v,(3)}_{L,-1} = - C^{v,(3)}_{-1}\,, && C^{a,(3)}_{L,-1,\text{diag}} = - C^{a,(3)}_{-1,\text{diag}}\,.
\end{align}
Note that in the non-diagonal case the vector and
axial-vector correlators conincide and that the ($-1$)-th moment of the diagonal
vector correlator vanishes. The contributions from W- and Z-boson self energies are divergent on their own and only
their sum is finite.  The singlet terms calculated in
Ref.~\cite{Schroder:2005db} are finite on their own and we do not repeat
them here. Using the results given in~\ref{sec:subtr} we obtain in the \MSbar{} scheme
\begin{equation}
  \label{eq:5}
  \bar{\rho}_{3,\text{non-singlet}} = \bar{C}^{a,(3)}_{-1,\mathrm{diag}} - \bar{C}^{v,(3)}_{-1} = 1.60667\,,
\end{equation}
and after converting to the on-shell scheme
\begin{equation}
  \label{eq:5OS}
  \rho_{3,\text{non-singlet}} = -101.083 \,,
\end{equation}
in full agreement with the results in the
literature~\cite{Chetyrkin:2006bj,Boughezal:2006xk}.

\section{Conclusion}
\label{sec:conclusion}

We have calculated the four-loop QCD corrections to the low-energy
moments of flavour non-diagonal current correlators up to $n=4$. Our
results are valid for (axial-)vector and (pseudo-)scalar currents in the
limit of a vanishing light-quark mass. As a by-product we have confirmed
the results for the non-singlet correction to the electroweak $\rho$
parameter first obtained in ~\cite{Chetyrkin:2006bj,Boughezal:2006xk}. In
combination with lattice simulations, our results can be used for the
precision determination of heavy-quark masses. Furthermore, they can
serve as an ingredient in the approximate reconstruction of the
four-loop corrections for arbitrary external momenta. For the latter
application, however, more input from other kinematic regions is
still required.

\section*{Acknowledgments}\label{sec:acknowledgements}
This work was supported by European Commission through contract
PITN-GA-2012-316704 (HIGGSTOOLS). We would like to thank
  Matthias Steinhauser for the reading of the manusript.

\appendix

\section{Symmetrisation}
\label{sec:symmetrisation}

The closely related problems of symmetrisation and mapping diagrams to
topologies are ubiquitous in multiloop calculations. Commonly used
algorithms employ either the diagrams' parametric
representations~\cite{Pak:2011xt} or representations as graphs. To avoid
cumbersome transformations, we choose to work with the original
algebraic form obtained directly from the Feynman rules.

A general $L$-loop scalar diagram $I$ with $P$ propagators has the form
\begin{equation}
  \label{eq:def_scalar_int}
  I = \int [dl_1] \dots [dl_L]\, \frac{1}{D_1^{a_1}\dots D_P^{a_P}}
\end{equation}
with (not necessarily positive) integers $a_1,\dots
a_P$. The $[dl_i]$ are suitable $d$-dimensional integral measures,
e.g. as in equation~(\ref{eq:def_d_int}), and the propagators $D_i$ are
functions of the loop momenta $l_1,\dots,l_L$, any number of external
momenta, and a mass $m_i$.
Obviously, $I$ is invariant under a change of variables
\begin{equation}
  \label{eq:shift_l}
  {\cal M}: l_i \mapsto l'_i = M_{ij}l_j + q_i
\end{equation}
with $|\det(M)| = 1$ and constant vectors $q_i$.

Consider now a diagram $\tilde{I}$ with propagators
$\tilde{D}_1,\dots,\tilde{D}_P$ and the diagram $I$ as defined by
eq.~(\ref{eq:def_scalar_int}). Let us denote the propagators we obtain
by changing the loop momenta in $I$ according to eq.~(\ref{eq:shift_l})
as $D'_1,\dots,D'_P$. We say that $I$ and $\tilde{I}$ belong to the
same topology iff there is a transformation ${\cal M}$ such that
$\{D'_1,\dots,D'_P\} = \{\tilde{D}_1,\dots,\tilde{D}_P\}$. Likewise,
$I$ belongs to a subtopology of $\tilde{I}$ iff for some ${\cal M}$ we
have $\{D'_1,\dots,D'_P\} \subseteq
\{\tilde{D}_1,\dots,\tilde{D}_P\}$. The problem of mapping a diagram to
a topology thus reduces to finding out whether a suitable transformation
${\cal M}$ exists.

The basic idea behind our algorithm is to first look for $L$ propagators
$D_i$ that depend on all loop momenta $l_1,\dots,l_L$. Then we select
$L$ appropriate mutually different target propagators $\tilde{D}_{j_i}$
and define ${\cal M}$ such that $D'_i = \tilde{D}_{j_i}$. If the sets of
the remaining propagators are also equal after applying ${\cal M}$, the
two topologies are the same.

To be more concrete, let us now consider a diagram $I$ defined as in
equation~(\ref{eq:def_scalar_int}) with propagators of the form $D_i =
p_i^2 \pm m_i^2$, where the $p_i$ are linear combinations of loop
momenta and external momenta. The generalisation to other forms of the
propagators should be straightforward. In practice, we can choose the
first $L$ propagators to be of the form $D_i = l_i^2 \pm m_i^2$. The
algorithm then works as follows.
\begin{enumerate}
\item Select a new target topology and choose a representative with
propagators $\{\tilde{D}_1,\dots,\tilde{D}_P\}$ of the form $\tilde{D}_i =
\tilde{p}^2_i \pm \tilde{m}^2_i$ from it.
\item Choose a tuple $(\tilde{D}_{i_1},\dots,\tilde{D}_{i_L})$ (that was
not chosen before) of $L$ distinct propagators with compatible masses,
  i.e. $\tilde{m}_{i_1} = m_1,\dots,\tilde{m}_{i_L} = m_L$. If this is
  not possible go back to step 1.
\item Consider the next among the $2^L$ transformations that map
  the propagators $(D_1,\dots,D_L)$ onto
  $(\tilde{D}_{i_1},\dots,\tilde{D}_{i_L})$, i.e. $l_j \mapsto \pm
  p_{i_j}\, j = 1,\dots,L$. If no transformation is left go back to
  step 2.
\item Apply the current transformation to the propagators
  $D_1,\dots,D_P$. $I$ then belongs to the current
  target topology if $\{D'_1,\dots,D'_P\} =
  \{\tilde{D}_1,\dots,\tilde{D}_P\}$. Else go back to step 3.
\end{enumerate}
As far as identifying the topology of an integral is concerned the
algorithm terminates as soon as step 4 is completed successfully. For
symmetrisation we would skip step 1 and always go back from step 4 to
step 3 in order to find all automorphisms.

\section{Subtraction terms}
\label{sec:subtr}

Since the leading coefficients with $n=-1, 0$ in
equation~(\ref{eq:C_def}) still depend on the dimensional regulator
$\epsilon = (4-d)/2$, we first have to specify our renormalisation
prescriptions in $d$ dimensions in order to give meaningful expressions.

Our $d$-dimensional integration measure is given by
\begin{equation}
  \label{eq:def_d_int} [dl_i] = \frac{d^dl_i}{i
\pi^{d/2}}e^{\epsilon\gamma_{E}}\,,
\end{equation} where $\gamma_{E} \approx 0.5772157$ is the
Euler-Mascheroni constant. The counterterms in the \MSbar{} scheme are
now defined such that they exactly cancel the poles in $\epsilon$. For
the sake of simplicity, we refrain from defining on-shell
renormalisation and present the divergent coefficients in terms of the
\MSbar{} quark mass. Writing
\begin{equation}
  \label{eq:c_def}
  \bar{C}^{\delta,(3)}_n = \sum_{i=0}^{3-n} \frac{\bar{c}^{\,\delta,(3)}_{n, i}}{\epsilon^i}
\end{equation}
we obtain for $\mu = \bar{m}$
\begin{align}
\label{cv_-1_0}
  \bar{c}^{\,v,(3)}_{-1,0} ={}&+1.740\,\*C_A^2\*C_F - 9.555\,\*C_A\*C_F^2 + 15.433\,\*C_F^3 \notag\\
& - 7.803\,\*C_A\*C_F\*T_F\*n_h + 7.355\,\*C_F^2\*T_F\*n_h \notag\\
& - 0.228\,\*C_A\*C_F\*T_F\*n_l - 1.897\,\*C_F^2\*T_F\*n_l \notag\\
& - 0.935\,\*C_F\*T_F^2\*n_h^2 + 0.735\,\*C_F\*T_F^2\*n_h\*n_l + 1.024\,\*C_F\*T_F^2\*n_l^2\,,
\displaybreak[0] \\
\label{cv_-1_1}
  \bar{c}^{\,v,(3)}_{-1,1} ={}&-1.196\,\*C_A^2\*C_F + 0.592\,\*C_A\*C_F^2 - 1.377\,\*C_F^3 \notag\\
& + 1.130\,\*C_A\*C_F\*T_F\*n_f + 0.015\,\*C_F^2\*T_F\*n_f + 0.009\,\*C_F\*T_F^2\*n_f^2\,,
\displaybreak[0] \\
\label{cv_-1_2}
  \bar{c}^{\,v,(3)}_{-1,2} ={}&+2.195\,\*C_A^2\*C_F + 0.649\,\*C_A\*C_F^2 + 1.278\,\*C_F^3 \notag\\
& - 1.623\,\*C_A\*C_F\*T_F\*n_f - 0.244\,\*C_F^2\*T_F\*n_f - 0.025\,\*C_F\*T_F^2\*n_f^2\,,
\displaybreak[0] \\
\label{cv_-1_3}
  \bar{c}^{\,v,(3)}_{-1,3} ={}&-1.058\,\*C_A^2\*C_F - 1.750\,\*C_A\*C_F^2 - 0.352\,\*C_F^3 \notag\\
& + 0.635\,\*C_A\*C_F\*T_F\*n_f + 0.531\,\*C_F^2\*T_F\*n_f  - 0.069\,\*C_F\*T_F^2\*n_f^2\,,
\displaybreak[0] \\
\label{cv_-1_4}
  \bar{c}^{\,v,(3)}_{-1,4} ={}&+0.210\,\*C_A^2\*C_F + 0.516\,\*C_A\*C_F^2 + 0.281\,\*C_F^3 \notag\\
& - 0.153\,\*C_A\*C_F\*T_F\*n_f - 0.188\,\*C_F^2\*T_F\*n_f  + 0.028\,\*C_F\*T_F^2\*n_f^2\,,
\displaybreak[0] \\
\label{cv_0_0}
  \bar{c}^{\,v,(3)}_{0,0} ={}&-0.832\,\*C_A^2\*C_F - 3.606\,\*C_A\*C_F^2 + 2.628\,\*C_F^3 \notag\\
& - 1.432\,\*C_A\*C_F\*T_F\*n_h + 2.335\,\*C_F^2\*T_F\*n_h \notag\\
& + 2.239\,\*C_A\*C_F\*T_F\*n_l + 0.666\,\*C_F^2\*T_F\*n_l \notag\\
& - 0.425\,\*C_F\*T_F^2\*n_h^2 - 0.479\,\*C_F\*T_F^2\*n_h\*n_l - 0.330\,\*C_F\*T_F^2\*n_l^2\,,
\displaybreak[0] \\
\label{cv_0_1}
  \bar{c}^{\,v,(3)}_{0,1} ={}&+0.277\,\*C_A^2\*C_F + 0.065\,\*C_A\*C_F^2 - 0.180\,\*C_F^3 \notag\\
& - 0.417\,\*C_A\*C_F\*T_F\*n_f + 0.172\,\*C_F^2\*T_F\*n_f  - 0.020\,\*C_F\*T_F^2\*n_f^2\,,
\displaybreak[0] \\
\label{cv_0_2}
  \bar{c}^{\,v,(3)}_{0,2} ={}&-0.230\,\*C_A^2\*C_F + 0.019\,\*C_A\*C_F^2 \notag\\
& + 0.150\,\*C_A\*C_F\*T_F\*n_f + 0.024\,\*C_F^2\*T_F\*n_f  - 0.017\,\*C_F\*T_F^2\*n_f^2\,,
\displaybreak[0] \\
\label{cv_0_3}
  \bar{c}^{\,v,(3)}_{0,3} ={}&+0.070\,\*C_A^2\*C_F - 0.051\,\*C_A\*C_F\*T_F\*n_f + 0.009\,\*C_F\*T_F^2\*n_f^2\,,
\displaybreak[0] \\
\label{cs_-1_0}
  \bar{c}^{\,s,(3)}_{-1,0} ={}&-72.707\,\*C_A^2\*C_F - 114.585\,\*C_A\*C_F^2 + 20.766\,\*C_F^3 \notag\\
& + 14.819\,\*C_A\*C_F\*T_F\*n_h + 101.776\,\*C_F^2\*T_F\*n_h \notag\\
& + 62.816\,\*C_A\*C_F\*T_F\*n_l + 19.095\,\*C_F^2\*T_F\*n_l \notag\\
& - 17.829\,\*C_F\*T_F^2\*n_h^2 - 24.041\,\*C_F\*T_F^2\*n_h\*n_l - 3.175\,\*C_F\*T_F^2\*n_l^2\,,
\displaybreak[0] \\
\label{cs_-1_1}
  \bar{c}^{\,s,(3)}_{-1,1} ={}&-5.959\,\*C_A^2\*C_F + 10.188\,\*C_A\*C_F^2 - 6.959\,\*C_F^3 \notag\\
& + 16.536\,\*C_A\*C_F\*T_F\*n_h -  0.578\,\*C_F^2\*T_F\*n_h \notag\\
& + 2.295\,\*C_A\*C_F\*T_F\*n_l +  0.422\,\*C_F^2\*T_F\*n_l \notag\\
& -  0.544\,\*C_F\*T_F^2\*n_h^2 - 0.644\,\*C_F\*T_F^2\*n_h\*n_l - 0.100\,\*C_F\*T_F^2\*n_l^2\,,
\displaybreak[0] \\
\label{cs_-1_2}
  \bar{c}^{\,s,(3)}_{-1,2} ={}&+7.939\,\*C_A^2\*C_F + 1.310\,\*C_A\*C_F^2 + 3.731\,\*C_F^3 \notag\\
& - 7.673\,\*C_A\*C_F\*T_F\*n_h -  3.327\,\*C_F^2\*T_F\*n_h \notag\\
& - 5.840\,\*C_A\*C_F\*T_F\*n_l - 0.327\,\*C_F^2\*T_F\*n_l \notag\\
& +  0.481\,\*C_F\*T_F^2\*n_h^2 + 0.296\,\*C_F\*T_F^2\*n_h\*n_l - 0.185\,\*C_F\*T_F^2\*n_l^2\,,
\displaybreak[0] \\
\label{cs_-1_3}
  \bar{c}^{\,s,(3)}_{-1,3} ={}&-3.813\,\*C_A^2\*C_F - 11.708\,\*C_A\*C_F^2 - 2.438\,\*C_F^3 \notag\\
& + 2.236\,\*C_A\*C_F\*T_F\*n_f + 3.042\,\*C_F^2\*T_F\*n_f  - 0.222\,\*C_F\*T_F^2\*n_f^2\,,
\displaybreak[0] \\
\label{cs_-1_4}
  \bar{c}^{\,s,(3)}_{-1,4} ={}&+0.840\,\*C_A^2\*C_F + 4.125\,\*C_A\*C_F^2 + 4.500\,\*C_F^3 \notag\\
& - 0.611\,\*C_A\*C_F\*T_F\*n_f - 1.500\,\*C_F^2\*T_F\*n_f  + 0.111\,\*C_F\*T_F^2\*n_f^2\,,
\displaybreak[0] \\
\label{cs_0_0}
  \bar{c}^{\,s,(3)}_{0,0} ={}&-1.740\,\*C_A^2\*C_F + 9.555\,\*C_A\*C_F^2 - 15.433\,\*C_F^3 \notag\\
& + 7.803\,\*C_A\*C_F\*T_F\*n_h - 7.355\,\*C_F^2\*T_F\*n_h \notag\\
& + 0.228\,\*C_A\*C_F\*T_F\*n_l + 1.897\,\*C_F^2\*T_F\*n_l \notag\\
& + 0.935\,\*C_F\*T_F^2\*n_h^2 - 0.735\,\*C_F\*T_F^2\*n_h\*n_l - 1.024\,\*C_F\*T_F^2\*n_l^2\,,
\displaybreak[0] \\
\label{cs_0_1}
  \bar{c}^{\,s,(3)}_{0,1} ={}&+1.196\,\*C_A^2\*C_F - 0.592\,\*C_A\*C_F^2 + 1.377\,\*C_F^3 \notag\\
& - 1.130\,\*C_A\*C_F\*T_F\*n_f - 0.015\,\*C_F^2\*T_F\*n_f - 0.009\,\*C_F\*T_F^2\*n_f^2\,,
\displaybreak[0] \\
\label{cs_0_2}
  \bar{c}^{\,s,(3)}_{0,2} ={}&-2.195\,\*C_A^2\*C_F - 0.649\,\*C_A\*C_F^2 - 1.278\,\*C_F^3 \notag\\
& + 1.623\,\*C_A\*C_F\*T_F\*n_f + 0.244\,\*C_F^2\*T_F\*n_f  + 0.025\,\*C_F\*T_F^2\*n_f^2\,,
\displaybreak[0] \\
\label{cs_0_3}
  \bar{c}^{\,s,(3)}_{0,3} ={}&+1.058\,\*C_A^2\*C_F + 1.750\,\*C_A\*C_F^2 + 0.352\,\*C_F^3 \notag\\
& - 0.635\,\*C_A\*C_F\*T_F\*n_f - 0.531\,\*C_F^2\*T_F\*n_f  + 0.069\,\*C_F\*T_F^2\*n_f^2\,,
\end{align}
with $n_f = n_h + n_l$.

In addition to the listed coefficient $\bar{C}^{\,v,(3)}_{-1}$ we require
the corresponding coefficient $\bar{C}^{\,a,(3)}_{-1, \text{diag}}$ in the
low-energy expansion of the flavour diagonal axial-vector correlator in
order to compute the correction to the $\rho$ parameter. Since the pole
parts of these two coefficients have to cancel, we can decompose the
latter coefficient as
\begin{equation}
  \label{eq:c_av_diag_decomp}
  \bar{C}^{a,(3)}_{-1, \text{diag}} = \bar{C}^{a,(3)}_{-1, \text{diag}}\bigg|_{\text{fin}} - \sum_{i=1}^{4} \frac{\bar{c}^{\,v,(3)}_{-1, i}}{\epsilon^i}
\end{equation}
with the coefficients $\bar{c}^{\,v,(3)}_{-1, i}$ as in
equations~\ref{cv_-1_1}--\ref{cv_-1_4}. The remaining finite part is
given by
\begin{align}
  \label{eq:c_av_diag_fin}
\bar{C}^{a,(3)}_{-1, \text{diag}}\bigg|_{\text{fin}} ={}&+2.484\,\*C_A^2\*C_F - 8.319\,\*C_A\*C_F^2 + 16.954\,\*C_F^3 \notag\\
& - 5.300\,\*C_A\*C_F\*T_F\*n_h +  2.759\,\*C_F^2\*T_F\*n_h \notag\\
& - 1.598\,\*C_A\*C_F\*T_F\*n_l - 4.210\,\*C_F^2\*T_F\*n_l \notag\\
& -  0.247\,\*C_F\*T_F^2\*n_h^2 + 1.585\,\*C_F\*T_F^2\*n_h\*n_l + 1.492\,\*C_F\*T_F^2\*n_l^2\,,
\end{align}

\section*{References}

\bibliographystyle{elsarticle-num}
 \bibliography{biblio}

\end{document}